\documentclass[twocolumn]{aa} % for a referee version
\usepackage{graphicx}
\usepackage{txfonts}
 \usepackage{lscape}
 \usepackage{colortbl}
 \usepackage{natbib}
 \usepackage{verbatim} 
\usepackage{url}
 \bibpunct{(}{)}{;}{a}{}{,}
\begin{document}
   \title{Evolution of blue E/S0 galaxies from $z\sim1$: merger remnants or disk rebuilding galaxies?}

  \author{M. Huertas-Company 
          \inst{1,2,3}
          \and
          J.A.L. Aguerri
          \inst{4}
          \and
          L. Tresse
          \inst{5}
          \and
          M. Bolzonella
          \inst{6}
          \and
          A. M. Koekemoer
          \inst{7}
          \and
          C. Maier
          \inst{8}
          }

   \institute{ESO, 
   		Alonso de Cordova 3107 - Casilla 19001 - Vitacura -Santiago, Chile	 \email{mhuertas@eso.org, marc.huertas@obspm.fr}
		\and
		GEPI - Observatoire de Paris, Section de Meudon, 5 Place Jules Janssen - 92190 - Meudon, France
		\and
		University of Paris Denis Diderot, 75205 Paris Cedex 13, France
		\and
		Instituto de Astrof\'isica de Canarias, C/ V\'ia L\'actea s/n, 38200 La Laguna, Spain
		\and
		LAM-CNRS Universit\'e de Provence, 38, rue Fr\'ed\'eric
Joliot-Curie, 13388 Marseille cedex 13, France 
                \and
                INAF - OA Bologna, via Ranzani 1, I-40127 Bologna - Italy
              \and
              Space Telescope Science Institute 3700 San Martin Drive, Baltimore MD 21218 - U.S.A.
                          \and
              Institute of Astronomy, Swiss Federal Institute of Technology (ETH
H\"onggerberg), CH-8093, Z\"urich, Switzerland}

   \date{}

% \abstract{}{}{}{}{} 
% 5 {} token are mandatory
 
  \abstract
   {Studying outliers from the bimodal distribution of galaxies in the color-mass space, such as morphological early-type galaxies residing in the blue cloud (\emph{blue E/S0s}), can help  to better understand the physical mechanisms that lead galaxy migrations in this space.}
   { In this paper we try to bring new clues by studying the evolution of the properties of a significant sample of blue E/S0 galaxies in the COSMOS field. }
   {We define blue E/S0 galaxies as objects having a clear early-type morphology on the HST/ACS images (according to our automated classification scheme \textsc{galSVM}) but with a blue rest-frame color (defined using the SED best fit template on the COSMOS primary photometric catalogues). Combining these two measurements with spectroscopic redshifts from the zCOSMOS 10k release, we isolate 210 $I_{AB}<22$ blue early-type galaxies with $M_*/M_\odot>10^{10}$ in three redshift bins ($0.2<z<0.55$, $0.55<z<0.8$, $0.8<z<1.4$) and study the evolution of their properties (number density, SFR, morphology, size).  }
   {The threshold mass ($M_t$) defined at z=0 in previous studies as the mass below which the population of blue early-type galaxies starts to be abundant relative to passive E/S0s, evolves from $log(M_*/M_\odot)\sim10.1\pm0.35$ at $z\sim0.3$ to $log(M_*/M_\odot)\sim10.9\pm0.35$ at $z\sim1$. Interestingly it follows the evolution of the crossover mass between the early and late type population (bimodality mass) indicating that the abundance of blue E/S0 is another measure of the downsizing effect in the build-up of the red-sequence. There seems to be a turn-over mass in the nature of blue E/S0 galaxies. Above $ log(M_*/M_\odot)\sim10.8$ blue E/S0 resemble to merger remnants probably migrating to the red-sequence in a time-scale of $\sim 3$ Gyr. Below this mass, they seem to be closer to normal late-type galaxies as if they were the result of minor mergers which triggered the central star-formation and built a central bulge component or were (re)building a disk from the surrounding gas in a much longer time-scale, suggesting that they are moving back or staying in the blue-cloud. This turn-over mass does not seem to evolve significantly from $z\sim1$ in contrast with the threshold mass and therefore does not seem to be linked with the relative abundance of blue E/S0s. }
   {}


   \maketitle
%
%________________________________________________________________

\section{Introduction}
It is well established that there  is a clear bimodality of the $z\sim0$ galaxy
distribution  in the  color-mass/magnitude  space. The  red  sequence (RS) is mostly
composed  by  red, passive  early-type  galaxies  and  the blue  cloud
contains blue, star-forming  late-type galaxies. One fundamental point
is to  understand the mechanisms  which drive the building-up  of this
relation.  What  are  the  movements  in this  color-mass  space?  How
galaxies are moving  from the blue-cloud to the  red-sequence and vice
versa?

 The blue cloud and the red sequence are not uniformly distributed in the
color-mass space. Red galaxies indeed dominate in number at large stellar masses and the crossover stellar mass between late and early-type
galaxies has been called  \emph{bimodality   mass}  ($M_{b}$) (e.g. \citealp{Strateva01, Baldry04, Kauffmann06, Faber07}). Recent
high-redshift studies  have shown that this bimodality mass has evolved from   $M_{b}\sim1-2\times10^{11}M_{\odot}$  at
$z\sim1$ to $M_b\sim3\times10^{10}$ at $z\sim0$ (e.g
\citealp{bundy05, cimatti06}).   This has been interpreted as another
signature  of  the downsizing  effect  \citep{Cowie96}  in the  galaxy
formation processes  with galaxies moving  from the blue-cloud  to the
red-sequence  at  higher  masses  when  we  look  back  in  time, even if, as argued by \cite{cattaneo08}, this interpretation is not always straightforward.  

While the \emph{basic} physics behind it has been known since a series of seminal papers published 20-30 years ago (e.g. \citealp{Rees77, Binney77, White78, blumenthal84}), a detailed understanding is still in progress, in particular through the works of \cite{dekel06} and \cite{Keres05}.  The result is that, in order to quench the star-formation in an efficient way and reproduce the mass-dependent transitions, recent semi-analitical models (e.g. \citealp{deLucia06, cattaneo06}) need to use different combinations of AGN feedbacks (e.g. \citealp{silk98,
  fabian99, king03,  wyithe03}) and merging (e.g.  \citealp{hernquist95,  mihos96,
  springel05}).
  %We need  processes to quench the star  formation above a mass
%threshold such  as gas loss  caused by AGN  feedback (e.g. \citealp{silk98,
  %fabian99, king03,  wyithe03}), consumption of all  available gas in
%the  wake  of  violent  merging (e.g.  \citealp{hernquist95,  mihos96,
 % springel05}), ram-pressure  stripping/harassment  in clusters  that
%turns off efficient  cold gas accretion (e.g. \citealp{quilis01, moore98}).
%Usually a combination  of  these  processes  are normally  required  in  semi
%analytical models  to ensure  that quenching is  efficient, permanent,
%and mass-dependent (e.g. \citealp{deLucia06}). 

From an observational point of view, it becomes interesting to study outliers in the color morphology  relation since they can represent objects in transition, and can
bring new clues  about its origin. As a matter of fact, some recent studies have pointed out, that at z=0, some low mass morphologically defined E/S0s start to appear in the blue cloud (e.g.  \citealp{Bamford09,
  kannappan09, Schawinski09}). These so called \emph{blue early type} galaxies have also been found at higher  redshift   (e.g.  \citealp{Ilbert06b, Ferreras09}). 
  
  %Their  intriguing mismatched  color-morphology could
%indicate a transitional state in color (e.g. a fading starburst galaxy
%after  a merger  which will  normally move  to the  red  sequence), in
%morphology (e.g. a disk-building system which is therefore moving back
%to the  blue-cloud) or a combination  of both (e.g.  a gas-rich merger
%remnant regenerating a young disk  from tidal debris). In other words,
%blue early-type  galaxies might be  moving from the blue-cloud  to the
%red-sequence, from  the red-sequence to  the blue-cloud or  within the
%blue-cloud.  Identifying and understanding  these movements  should give
%new clues about the building-up of the morphology-mass bimodality. 

However, the detailed analysis of blue E/S0 galaxies has just started. At $z=0$,
\cite{kannappan09} found  a clear dependence on mass of  the properties of these objects. They basically identified three
mass  regimes:  above   the  shutdown  mass  $M_s\sim10^{11}M_{\odot}$
blue-early type galaxies are non-existent (as most of the blue-cloud),
below  the  threshold  mass  ($M_t\sim10^{9}$) they  become  extremely
abundant, representing  $\sim20-30\%$ of  all the E/S0  population and
between $M_s$ and  $M_t$ where the bimodality mass  lies, they account
for  $\sim5\%$ of  the E/S0  population. The  nature of  the physical
processes taking place in these mass regimes seem to be different as
well. As a matter of fact, the shutdown  mass could be related with the different
temperature of the gas accreted by dark matter haloes above and below a
critical  mass  of  $\sim  10^{12}M_{\odot}$  (e.g.  \citealp{bundy05,
  cattaneo06, dekel06}).   Blue early-type galaxies in  the mass range
$M_{s}<M<M_{b}$ resemble to merger remnants that are probably moving
to the red-sequence \citep{kannappan09}. Hierarchical  formation  theories predict indeed that  most  of the  spiral
galaxies have undergone a major  merger event in  the last 8  Gyr. Those
mergers between galaxies with a large fraction of gas provoke a short
(0.1 Gyr) and strong peak of star formation (e.g. \citealp{springel05,
  springel05b}). Mergers do not  only change the stellar content of
the   galaxies  but   also  induce   a  dramatic   change   of  their
morphology. In particular, the  disks are suppressed producing a compact
morphology  that   could  be   associated with an early-type  galaxy
\citep{barnes02}. In contrast, below $M_t$ blue early-type galaxies are
likely  rebuilding disks  and therefore  moving to  or staying  in the
blue-cloud  \citep{kannappan09}. Major  mergers also induce shock-heated
gas winds as well as the formation of  a remnant gas halo with a large cooling
time and could produce an enriched  medium with which a gas disk could
be formed  \citep{cox04}.  Recently, some examples  of rebuilding disks have also been  found in  luminous  compact  blue
galaxies \citep{puech07, hammer09}. It is  interesting  that $M_{t}$
also corresponds  with the mass  below which galaxies have  larger gas
reservoirs \citep{kannappan04, kannappan08}. 

The next step in the characterization of blue E/S0s is to follow them at higher redshift. As a matter of fact since the transition from the blue-cloud to the red-sequence is mass-dependent, we expect these different  characteristic masses to evolve with look-back time and therefore the properties of blue E/S0s for a given mass as well.  

In  this work we look for blue E/S0 galaxies, similar to those
reported in recent studies in the nearby Universe, between $z\sim0.2$ and $z\sim1.4$ in the spectroscopic follow-up 
of the galaxies detected in the HST/ACS COSMOS field (ie. the public 
10k-bright sample, \citealp{Lilly09}). We investigate their observational properties (number density, morphology, size, star-forming rate), compare them with  similar galaxies at z=0  \citep{kannappan09, Schawinski09} and try to give new clues about their nature and evolution.  
The paper proceeds as
follows: in  \S~\ref{sec:data} and \S~\ref{sec:phys_prop}  we describe
our  sample and  the  methods  to obtain  physical  properties of  the
galaxies.  \S~\ref{sec:blueE}   is  focused  on   the  definition  and
characterization of blue early-type galaxies at different redshifts. Finally, in \S~\ref{sec:discussion} we discuss the main results and conclusions are given in ~\S~\ref{sec:conclusions}. 

Throughout the paper, we assume that the Hubble constant, the matter density, and the
cosmological constant are $H_0=70$ km s$^{-1}$ Mpc$^{-1}$, $\Omega_m=0.3$ and $\Omega_\Lambda=0.7$ respectively. All magnitudes are in the AB system.

\section{Dataset and sample selection}
\label{sec:data}
%\subsection{Samples and galaxy selection}

We use  the public available  zCOSMOS 10k sample corresponding  to the
second  release  of  the  zCOSMOS-bright survey  \citep{Lilly09}.  The
zCOSMOS-bright,  aims to  produce a  redshift survey  of approximately
20,000 I-band selected  galaxies ($15<I_{AB}<22.5$). Covering the
approximately  1.7 $deg^2$  of  the COSMOS  field \citep{Scoville07}  (essentially the  full ACS-covered  area),  the  transverse  dimension  at $z  \sim1$  is  75
Mpc.  This   second  release  (DR2)   contains  the  results   of  the
zCOSMOS-bright spectroscopic observations that were carried out in VLT
service mode during the period April  2005 to June 2006. 83 masks were
observed, yielding 10643 spectra of galaxies with $IAB<22.5$. The data
were  reduced  by  the  zCOSMOS  team  and  prepared  for  release  in
collaboration  with ESO  (External Data  Products  group/Data Products
department). 

We cross correlate this  catalog with a morphological catalog obtained
on ACS images up to  $I_{AB}<22$ in order to have both the spectroscopic redshifts
and  the morphological  information. Details  on how  morphologies are
obtained  are  described  on  \S~\ref{sec:morpho}.  This  ACS  I-band
catalog is  part of the COSMOS HST/ACS  field \citep{Koekemoer07}. The
data set  consists of a  contiguous 1.64 $deg^{2}$ field  covering the
entire COSMOS  field. The Advanced  Camera for Surveys  (ACS) together
with the F814W filter (``Broad I'') were employed. 

The  final catalogue  contains 6240  galaxies with  magnitudes ranging
from $18<I_{AB}<22$.  Since the zCOSMOS-bright catalogue  is not entirely
released yet,  the selected sample does  not have all  the galaxies in
this  magnitude  range.  This  is   not  critical  if  the  sample  is
representative of  all the galaxy  population and no  selection biases
are introduced for different magnitudes.  To check this point, we look
at  the fraction  of  all  the $I_{AB}<22$  magnitude  limited sample  (ACS
imaging)  which are  selected  in  our sample. The  selected sample represents  $\sim20\%$ of
the whole sample  and the selection function is  basically flat at all
magnitudes and  for all morphologies.  We therefore consider  that the
sample is representative of the galaxy population and unbiased.  

Since  we will  analyze the  sample properties  as a  function  of the
stellar  mass, in  ~\ref{sec:stellar_mass}  we will  discuss the  mass
completeness in detail.

In this work  we study galaxies in three  redshift bins ($0.2<z<0.55$,
$0.55<z<0.8$, $0.8<z<1.4$).  These three  bins have similar  number of
objects within a factor 2.

%\begin{figure} 
%\centering 
 %\resizebox{\hsize}{!}{\includegraphics{complete_mag.ps}}
 %\caption{Selection function: fraction of galaxies in our selected sample sample as a function of IAB apparent magnitude. The blue line are late-type galaxies and the red line are early-type galaxies. } 
 %\label{fig:compl}
 %\end{figure}

\section{Physical quantities}
\label{sec:phys_prop}
\subsection{Morphologies}
\label{sec:morpho}
The high angular resolution measured on HST images makes them ideal to
properly estimate galaxy morphologies.  Galaxies at  low  and  high
redshift   have  been  historically   classified  by   parametric  and
non-parametric methods.  Among  the parametric methods for classifying
galaxies   we  can  mention   the  bulge-disc   decomposition  method,
consisting on a 2-D  fit  to the  surface-brightness  distribution of  the
galaxies (e.g. \citealp{trujillo01,
  aguerri02,   peng02,  simard02,  trujillo04,   aguerri04,  dejong04,
  aguerri05, mendezabreu08}).  On the other side, the non-parametric methods
classify   galaxies  without  any   assumption on the  different
components of  the galaxies. In  this case the classification  is based on some global properties of galaxies, such as
color, asymmetry,  or light concentration  \citep{abraham94, abraham96,
  vandenbergh96}.  In our  case,  morphologies are  determined in  the
I-band  ACS images  using  our own developed code \textsc{galSVM}\footnote{\url{http://www.lesia.obspm.fr/~huertas/galsvm.html}}. This non-parametric
N-dimensional code is based on support  vector machines (SVM) and uses a
training    set    built   from    a    local   visually    classified
sample. It has been tested and validated in \citep{huertas-Company08}.   Since   in   this   particular
application the spatial  resolution of the data is  high, the training
is  built  directly from the   real  data  by  performing  a  visual
classification on  a randomly selected subsample. More  details on how
this   is   performed  can   be   consulted   in \cite{tasca09} and
\cite{huertas-Company09}. 

Basically we separate galaxies  in two broad morphological types (late
and early-type). By late-type  we mean spiral and  irregular galaxies
and early-type galaxies include  elliptical and lenticular types.  The
code gives  for each  galaxy a probability  measure of belonging  to a
given morphological class. Since the classification is fully automated, we expect some errors in the classification. A visual inspection of most of the objects up to $z\sim1$ reveals that, while the majority of the galaxies are indeed E/S0s there are some errors which increase at higher z. Most of this misclassified objects are irregular or merger galaxies with a strong starbust that produces a high concentration of light. As shown in \cite{huertas-Company09}, the
probability  parameter is  a good  measure of  the reliability  of the
classification; high probabilities indicating that the classification is
reliable. In our case, only $\sim5\%$ of the galaxies have associated \emph{ambiguous} probability values between 0.4 and 0.6, so we believe our classification to be robust despite of a few misclassifications. 

In  the  following,  we  decide  that  a galaxy  belongs  to  a  given
morphological   class  whenever   the  probability   is   bigger  than
$p=0.5$. However, in order to  check the robustness of our results and
the  dependence on  possible miss  classifications, we  will sometimes
increase the probability threshold.

\begin{comment}
\subsection{Spectroscopic redshifts}

Spectroscopic    redshifts   are    computed    using   the  publicly available zCOSMOS
spectra\footnote{\url{http://archive.eso.org/archive/adp/zCOSMOS/VIMOS_spectroscopy_v1.0}}.  End-to-end data reduction  is carried  out using  the VIPIGI
software package \citep{Scodeggio05}.  Determination of redshifts is a
multi-step  process  and  involves  the use  of  different  approaches
tailored   to  the   individual   spectra.  These   include  first   a
computer-aided determination based  on cross-correlation with template
spectra coupled to continuum fitting and principal component analysis,
using  the  EZ  software   (Garilli  et  al.,  in  preparation).  This
preliminary  automated   step  is   followed  by  a   detailed  visual
examination  of the  one-  and two-dimensional  spectrograms of  every
object to critically assess the validity of the automated redshift. In
those cases  where the  automatic procedure fails,  a new  redshift is
computed based  on the wavelengths  of recognized features.  Two fully
independent reductions are carried  out of each spectrum, yielding two
independent redshift measurements. These are compared and ``reconciled"
(generally in  a face-to-face meeting)  to yield a final  redshift and
confidence class. 

In this work  we study galaxies in three  redshift bins ($0.2<z<0.55$,
$0.55<z<0.8$, $0.8<z<1.4$).  These three  bins have similar  number of
objects within a factor 2.
\end{comment}

\subsection{Absolute magnitudes and rest-frame color classification} 

We restricted the above morpho-spectroscopic
$0.2<z<1.4$ catalogue to redshifts consistent with the photometric
redshifts (Ilbert et al. 2009) to a precision of $0.1(1+z)$. This
selection excludes galaxies with reliable redshifts but with a failed
photometric redshift due plausible photometric problems. This final 
selection results in a catalogue of 6064 galaxies.  Absolute
(AB) magnitudes are derived using the cross-correlated spectro-photometric
catalogue including the following bands; NUV-2800$\AA$ GALEX \citep{Zamojski07}, $u^*$ CFHT/MegaCam and $BVg^{'}r^{'}i^{'}z^{'}$ SUBARU/SUPrime-Cam \citep{Capak07}, $K_s$ CFHT/WIRCam (McCracken et al., 2009), 3.6$\mu$m and 4.5$\mu$m SPITZER/IRAC \citep{Sanders07}. They are optimized in a way which minimizes the dependency
on the templates used to fit the multi-band photometry (see e.g. Fig
A.1. in Ilbert et al. 2005). We use templates generated with the
galaxy evolution model PEGASE.2 (Fioc and Rocca-Volmerabge, 1997), and
the multi-band input photometry is optimized as described in \cite{Bolzonella09}.

We aim at classifying galaxies according their rest-frame colors from
red to blue, and not according their PEGASE galaxy type. We applied
the same technique as described in Zucca et al. (2006) with 38
templates. For this work, we consider the red galaxy
subsample (templates [T1-T20]) and the blue one (templates [T21-T38]).

\begin{comment}
We use SED fitting with the primary photometric COSMOS catalogues to estimate rest-frame absolute magnitudes: the
NUV-2800$\AA$ GALEX near-ultraviolet catalogue \citep{Zamojski07}, the
$u^{*}$     CFHT/MegaCam    optical     catalogue,     the    $BVgriz$
SUBARU/SUPrime-Cam   optical   catalogue   \citep{Capak07},   the   Ks
CFHT/WIRCam near-infrared  catalogue (McCracken et al., in prep.), the  3.6$\mu$m and
4.5$\mu$m           SPITZER/IRAC           infrared          catalogue
\citep{Sanders07}.  Photometric  catalogues  have  been  cross-matched
within  a radius  of 1  arcsec with  the  spectroscopic zCOSMOS-Bright
selected sources. 

The sample  is restricted  to spectroscopic redshifts  consistent with
the   photometric  redshift   \citep{ilbert09}  to   a   precision  of
$0.1(1+z)$.   This  way,  the   derived  intrinsic   luminosities  and
rest-frame colors are ensured to be correct. This selection results in a catalog of 6064 galaxies. 

We associate a rest-frame color type using a best fitting technique as
done  for  the  VVDS  \citep{Zucca06}.  We  reduce  the  diversity  of
templates by  taking a subsample of  the templates used  to derive the
absolute  magnitudes.  The aim  is  to  classify  galaxies with  their
rest-frame colors instead of with the galaxy type. We therefore use 38
templates that we  regroup in two major rest-frame  for this work: the
red  templates [T1-T20]  and the  blue  ones [T21-T38].  
\end{comment}

\subsection{Stellar masses}
\label{sec:stellar_mass}
\begin{figure} 
 \centering 
  \resizebox{\hsize}{!}{\includegraphics{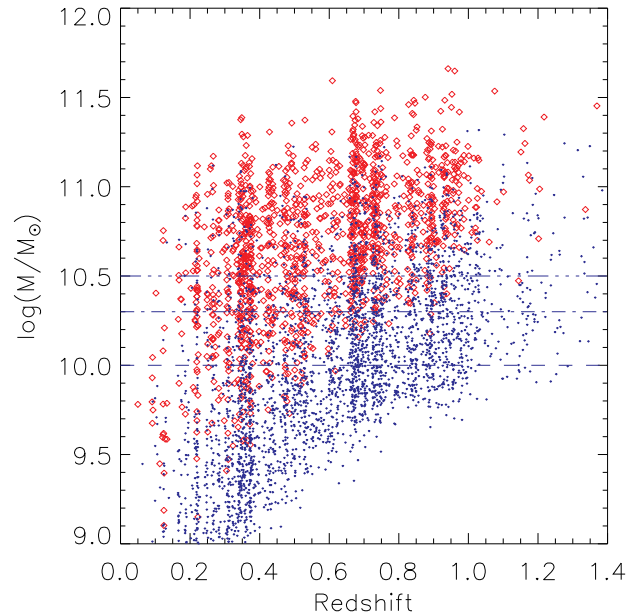}}
 \caption{Stellar mass vs. redshift. Blue points are blue late-type galaxies and red diamonds are red early-type galaxies. Horizontal lines from bottom to top show the estimated mass completeness at $z<0.55$, $z<0.8$ and $z<1.4$ respectively.} 
 \label{fig:mass_z}
 \end{figure}

Stellar masses are estimated from the best fitting template. In
this paper, we use the estimates performed in \cite{Bolzonella09}.
Since we  are trying  to identify  trends with the  stellar mass  on a
magnitude  limited sample, it  is crucial  to properly  understand the
completeness effects. As  a matter of fact, at  the highest redshifts,
the reddest  objects will begin to  drop out of the  sample. Figure~\ref{fig:mass_z}
shows the redshift of  the selected  objects as a function  of their
stellar  masses. It  is  evident  that less  massive objects are not
observed at high redshift, while the most massive ones are observed at
all redshifts. Notice also that  the blue and red galaxies have
different completeness limits: blue galaxies can be observed at higher
redshift than the red ones for a fixed mass. We therefore establish that our
completeness limiting mass of the total sample is $\sim 5\times10^{10}$ $M_*/M_\odot$ (as the mass above which red galaxies are still detected at $z\sim1$). All galaxies with masses
 larger than the completeness limit can be observed at all redshifts. We
can also compute the completeness mass for each of the three redshift
bins, being: $\sim10^{10}$ $M_*/M_\odot$, $\sim 10^{10.3}$ $M_*/M_\odot$  and $\sim 10^{10.5}$ $M_*/M_\odot$ for  galaxies located at  $z<0.55$,
$z<0.8$ and $z<1.4$, respectively.

%To translate the I-band cut into a conservative estimate this completeness mass, we follow the\cite{Fontana03} and \cite{Bundy05} way of doing. We consider for that purpose the likelihood of observing mock galaxies of known mass and color based on the Ks-band limiting magnitude. We therefore track the stellar mass of a template galaxy (PEGASE code, \citealp{Rocca}) with a reasonable $M_{*}/L$ ratio determined by models with solar metallicity and a burst of star formation starting at $z_{form}=10$. We then set the luminosity to the I-band limiting magnitude ($I_{AB}=22$) and the maximum stellar mass of these models provide estimates of completeness. We find that the mass completeness limit rises from $\sim9.10^{9}M_\odot$ at $z\sim0.3$, $\sim4.10^{10}M_\odot$ at $z\sim0.7$ and $\sim8.10^{10}M_\odot$ at $z\sim0.9$. 

\section{Main properties of blue sequence E/S0 galaxies}
\label{sec:blueE}
\subsection{Definition}
\label{sec:def}
\begin{figure*} 
 \centering 
  {\includegraphics[width=17cm]{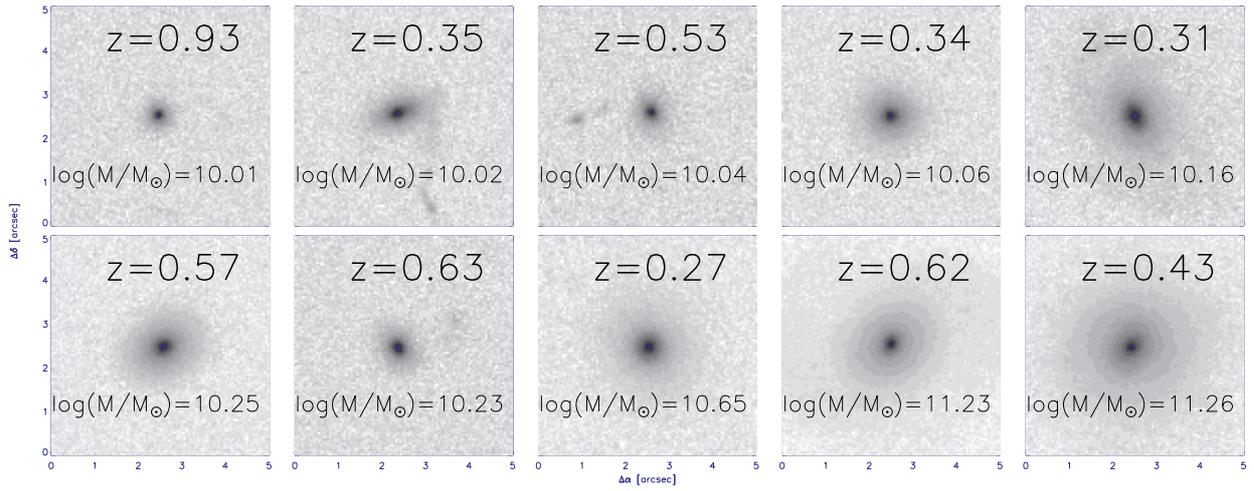}}
  \caption{$5^{"}\times5^{"}$ I-band ACS stamps of 10 blue E/S0 galaxies in our sample. Galaxies are sorted with increasing stellar mass. For each galaxy we indicate its spectroscopic redshift and its stellar mass. } 
 \label{fig:blueE_stamps}
 \end{figure*}

\begin{figure*} 
 \centering 
  {\includegraphics[width=17cm]{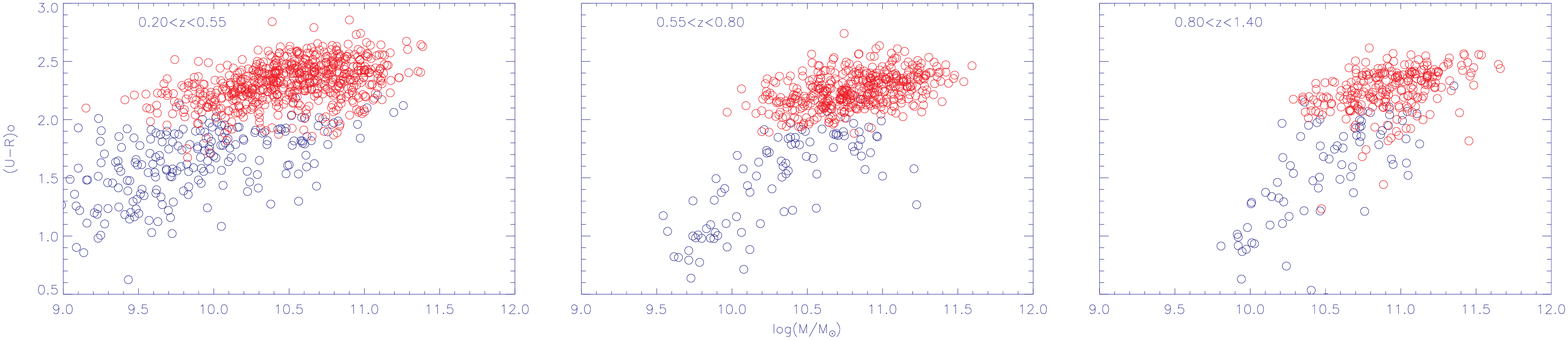}}
   {\includegraphics[width=17cm]{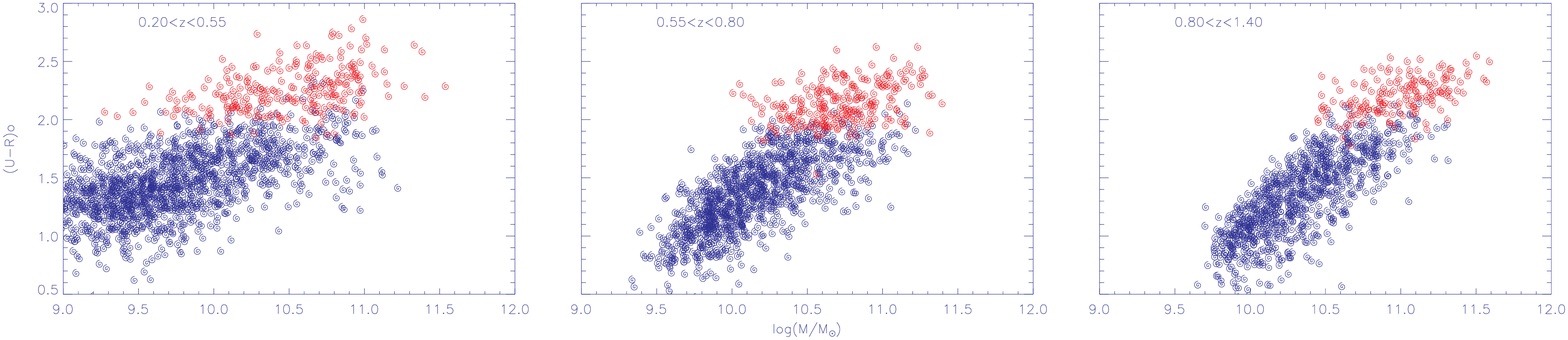}}
 \caption{(U-R) rest-frame color -- stellar mass diagrams for morphologically selected early-type galaxies (top) and late-type galaxies (down) in three redshift bins. Colors indicate the rest-frame color based classification: objects with red templates ([T1-T20]) are shown in red and objects with blue templates ([T21-T38]) are shown in blue. } 
 \label{fig:color_mass}
 \end{figure*}

We  define blue  early-type galaxies  as galaxies  with  an elliptical
morphology  ($p>0.5$) but  which SED  best fits  with a  blue template
([T21-T38]). As illustration, we show in figure~\ref{fig:blueE_stamps} some ACS/HST I-band stamps of these galaxies. The majority of them appear to be symmetric and with light concentrated towards the center of the galaxy which explains why they are classified as early-type. Their SED indicates however an on-going star formation. Figure~\ref{fig:color_mass} indeed shows the mass  color diagram for late and  early-type galaxies. Notice that the  blue E/S0 galaxies
are located in the so-called blue cloud at the same position as normal
blue spiral galaxies.

 %\begin{figure} 
 %\centering 
  %\resizebox{\hsize}{!}{\includegraphics{blueE_fraction.ps}}
 %\caption{Fraction of galaxies in our selected sample sample as a function of IAB apparent magnitude. The blue line are late-type galaxies and the red line are early-type galaxies. } 
 %\label{fig:blue_frac}
 %\end{figure}

 \subsection{Abundance}
 \label{sec:numbers}
\begin{figure} 
 \centering 
  \resizebox{\hsize}{!}{\includegraphics{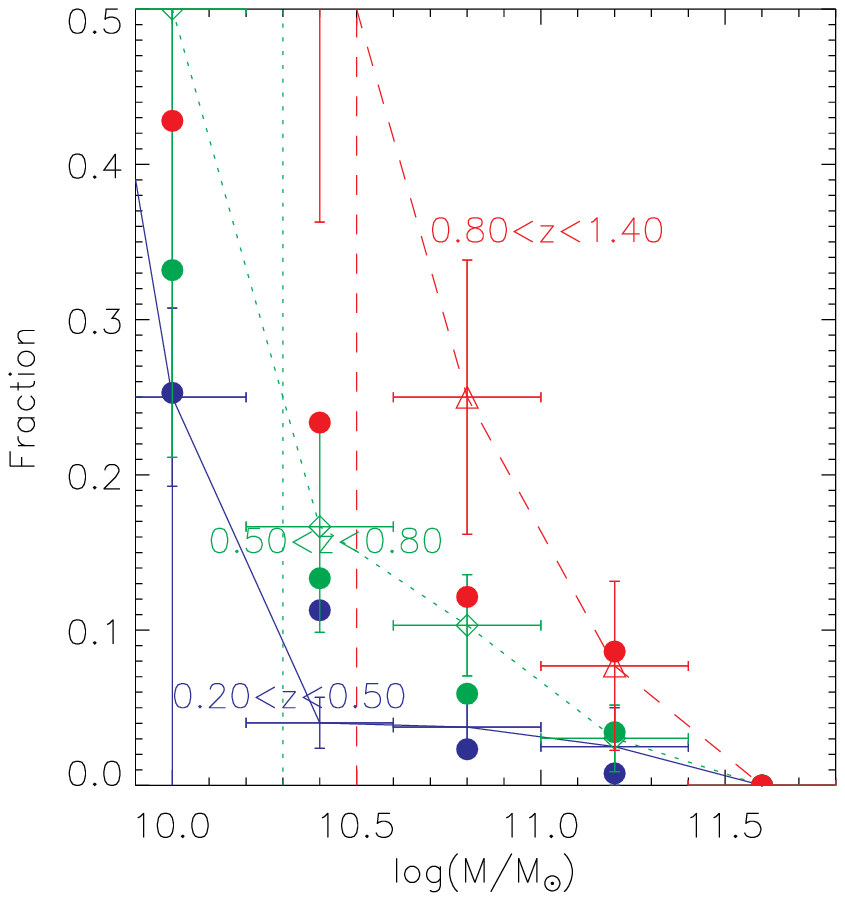}}
 \caption{Fraction of blue early-type galaxies as a function of mass at different redshifts. Red line: $0.8<z<1.4$, green line: $0.5<z<0.8$ and blue line:  $0.2<z<0.5$. Vertical lines show the estimated mass completeness at the different redshift bins. Filled circles show the results from the Millenium simulation at all redshifts (see text for details)} 
 \label{fig:blueE_single}
 \end{figure}

 From this section, we just  consider galaxies with stellar masses above the estimated mass
completeness    limit    at    $z\sim0.2$,    i.e.    galaxies    with
$log(M_*/M_\odot)>10$ to  avoid effects related to  the incompleteness of
the sample. Using this threshold we might have some incompleteness effects in the higher redshift bins but we will properly take this into account when required. Above this mass threshold, there are 3864 galaxies between
$z=0.2$ and $z=1.4$. Blue E/S0s represent $\sim6\%$ of
the whole  sample (210 galaxies).  The global morphological  mixing is
$\sim46\%$  of blue late-type  galaxies, $\sim33\%$  of red  E/S0s and
$\sim15\%$ red late-type galaxies which most of them are edge-on late-type systems
obscured by dust (visual inspection). Globally blue early-type galaxies therefore represent a small fraction of the galaxy distribution between $z\sim0.2$ and $z\sim1.4$. However, the distribution of these objects is not the same at all masses. \\
 \cite{kannappan09} defined a threshold  mass ($M_t$) at $z\sim0$ below
which  the  blue  early-type  population  starts  to  become  abundant
($\sim20\%-30\%$ of  the total E/S0 population). They  showed that the
nature  of  blue E/S0  galaxies  is  different  above and  below  this
threshold.   It  appears   therefore  interesting   to   quantify  this
characteristic mass at high redshift to see if we detect any evolution
and   if    it   is    still   linked   to    the   nature    of   the
galaxies.  Figure~\ref{fig:blueE_single}, shows  the evolution  of the
fraction  of  blue ellipticals  as  a  function  of stellar  mass  for
different redshifts.  At all  redshifts we do  see an increase  of the
abundance of this population when  the stellar mass decreases. We  decide  to perform  a rough  and  conservative
estimate of  the threshold  mass as the  mass at which the blue early-type
fraction becomes  higher than $20\%$.  Table~\ref{tbl:mass_evol} shows
the  measured values.  The  value rises  from $M_t\sim10.10\pm0.35$  at
$z\sim0.3$ to $M_t\sim10.90\pm0.35$  at $z\sim1$. We must nevertheless 
be   extremely  careful  in   analyzing  these   trends  because   of  mass
uncompleteness  effects. As  a matter  of fact,  due to  our magnitude
limited selection,  red objects  are the first  in falling out  of our
sample.  Therefore,  at low  masses  the  fraction  of red  early-type
objects might  be under-estimated, causing an  artificial increase of
the  blue  fraction.   Since our estimates
of $M_t$ is at all redshift bins greater than the completeness mass of
the  corresponding  bin, we  do  not  expect  our measurements  to  be
strongly affected by incompleteness. We also show in~\ref{fig:blueE_single} the results from the Millennium simulations \citep{springel05} and more precisely from the \emph{Durham} semi-analytical model \citep{bower06, benson03, cole00}. Blue E/S0s galaxies are selected as galaxies having blue colors ($U-R<1.1$) and a bulge stellar mass greater than $80\%$ of the total stellar mass. The color threshold is selected as the value which best divides the two bimodal peaks of the color histogram. This color threshold is different from the value one would suggest by looking at figure~\ref{fig:color_mass} which is closer to $\sim2$. This discrepancy might be explained by a difference in the magnitude system used (AB vs. Vega) or the filters sets employed. Despite of the fact that the selection criteria are not exactly the same than in our observational data, the abundance of these objects in the simulation is consistent with our measures which confirms our morphological classification and supports the fact that, at the considered stellar masses, we are not affected by incompleteness. 

\begin{figure*} 
 \centering 
  {\includegraphics[width=17cm]{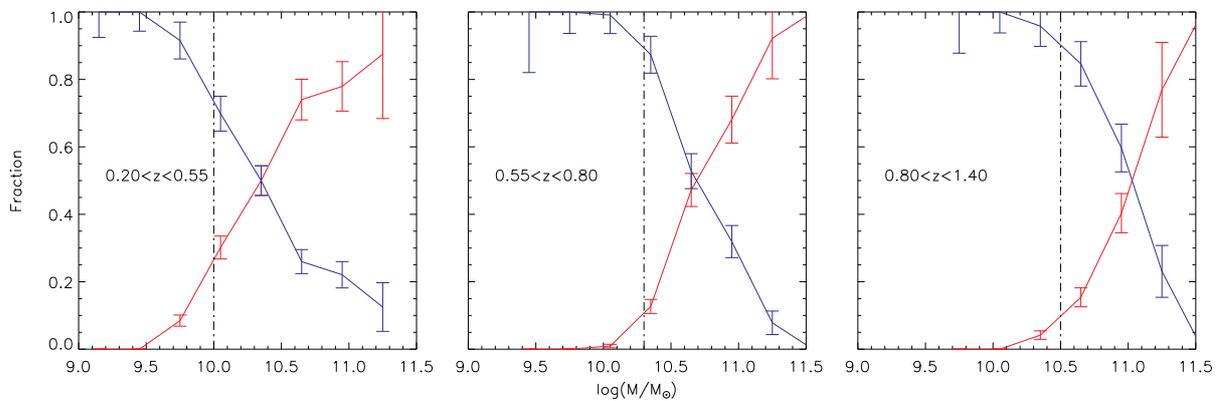}}
 \caption{Fractions of blue late-type (blue line) and red early-type galaxies (red line) in the three considered redshift bins as a function of the stellar mass. Dashed vertical lines show the estimated mass completeness at the corresponding redshift bin. } 
 \label{fig:bimo_mass}
 \end{figure*}

Another interesting mass is the bimodality mass ($M_b$) defined as the
crossover  mass between  the early-type  and the  late-type population
(e.g. \citealp{Kauffmann06}).  This mass is in  principle not directly
related  to the  blue  E/S0 population  but  it is  a  measure of  the
building-up of  the RS and could  therefore give clues on  the role of
blue E/S0  galaxies in this  process. Above this mass,  the early-type
population dominates  in number over  the late-type one.  Several high
redshift studies (e.g. \citealp{bundy05,pozzetti09}) have shown that this mass is
evolving with look-back time becoming higher at higher redshifts. This
evolution is interpreted as a  signature of the downsizing assembly of
galaxies.  Figure~\ref{fig:bimo_mass}  shows  the  fractions  of  both
morphological  types  in  the  three  considered redshift  bins  as  a
function  of the  stellar mass.  At all  redshits, we  do see  a clear
increase in  number of the  early-type population when  the considered
stellar mass  is higher. Notice  that at all redshifts,  the turn-over
mass  is beyond  the mass  completeness  limit, which  means that  the
estimate is not affected by possible uncompleteness effects. In order to properly estimate  the cross-over  mass  we perform  a  linear  fit to  the
declining  fraction of late-type  galaxies and  estimate $M_b$  as the
mass  where  the  best  fine   line  equals  0.5.  We  find  this  way
$log(M_b)=10.37\pm0.1$  at  $0.2<z<0.55$, $log(M_b)=10.72\pm0.08$  at
$0.55<z<0.8$ and $log(M_b)=11.00\pm0.2$  at $0.8<z<1.4$. To check the
robustness  of  these  estimates  with respect  to  the  morphological
classification,   we  repeated   the   measurement  using   increasing
probability thresholds.  The fluctuations of the  estimated values are
within  the  error bars.  We  therefore  confirm  the expected  trend,
i.e.  the  bimodality mass  increases  by  1  dex from  $z\sim0.2$  to
$z\sim1$.  This reflects the fact that the red-sequence  is first  populated with
massive galaxies.

The last interesting mass is a limitting mass  above  which no  blue galaxies  are
observed.  This mass was  defined by \cite{kannappan09} as the
shutdown-mass  ($M_s$). Figure~\ref{fig:color_mass} seem to show  that this  mass for blue E/S0s is close to $\sim 3
\times 10^{11} M_{\odot}$ at all redshifts.

% {\bf In order to quantify it in a more robust way, we computed the 95\% percentile of the distribution of blue E/S0s and all blue galaxies in our sample. }

\begin{table*}
\caption{Characteristic mass for different redshift bins. See text for more details on how these characteristic masses are defined. Results at $z\sim0$ are taken from \cite{kannappan09}}. 
\begin{tabular}{c|c|c|c|c}
\hline\hline\noalign{\smallskip}
&$z\sim0$  &$0.2<z<0.55$ & $0.55<z<0.8$ &  $0.8<z<1.4$ \\ 
\noalign{\smallskip}\hline
$log(M_s/M_\odot) $ &  11.20  &$11.50\pm0.38$ & $11.50\pm0.58$ & $11.50\pm1.20$ \\
\noalign{\smallskip}\hline
$log(M_b/M_\odot)$ &  10.50  &$10.37\pm0.1$ & $10.72\pm0.08$ & $11.00\pm0.2$ \\
\noalign{\smallskip}\hline
$log(M_t/M_\odot)$ & 9.70 & $10.10\pm0.35$ & $10.30\pm0.35$ & $10.90\pm0.35$ \\
\label{tbl:mass_evol}
\end{tabular}
\end{table*}

\subsection{Morphology}
\label{sec:morpho_stamps}
\begin{figure*} 
 \centering 
  {\includegraphics[width=17cm]{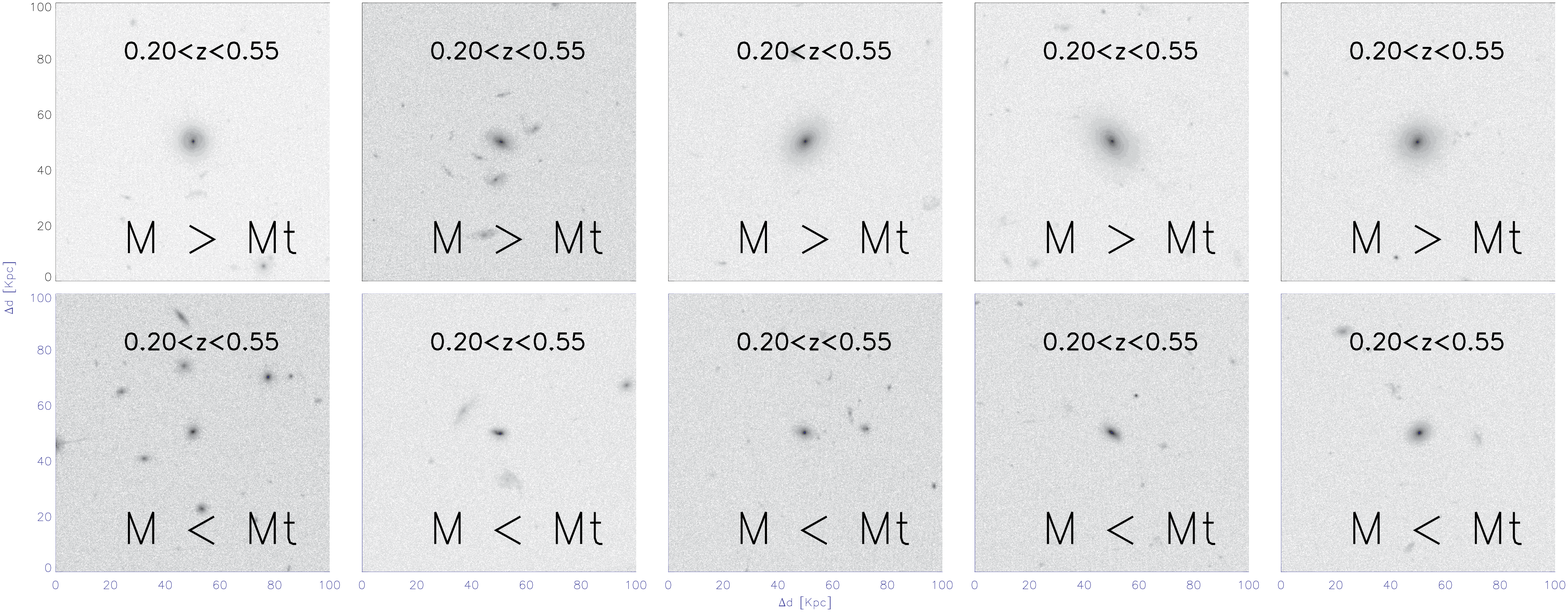}}
   {\includegraphics[width=17cm]{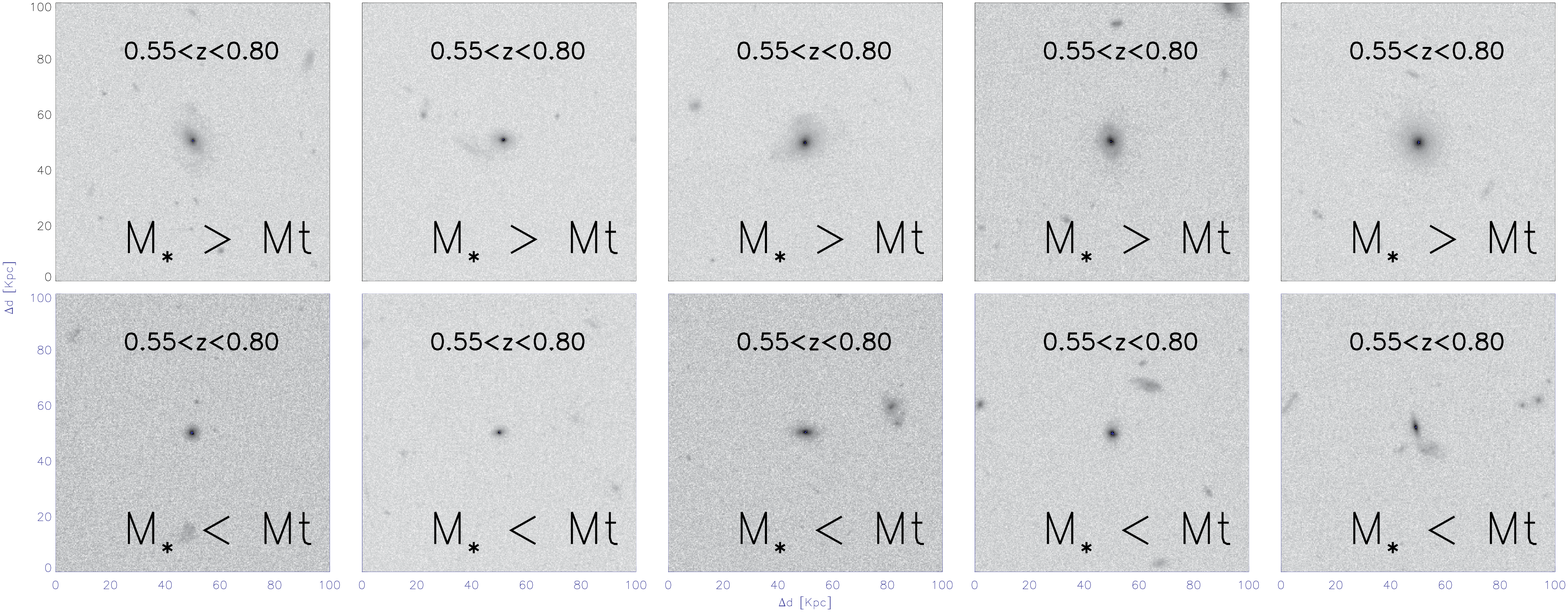}}
    {\includegraphics[width=17cm]{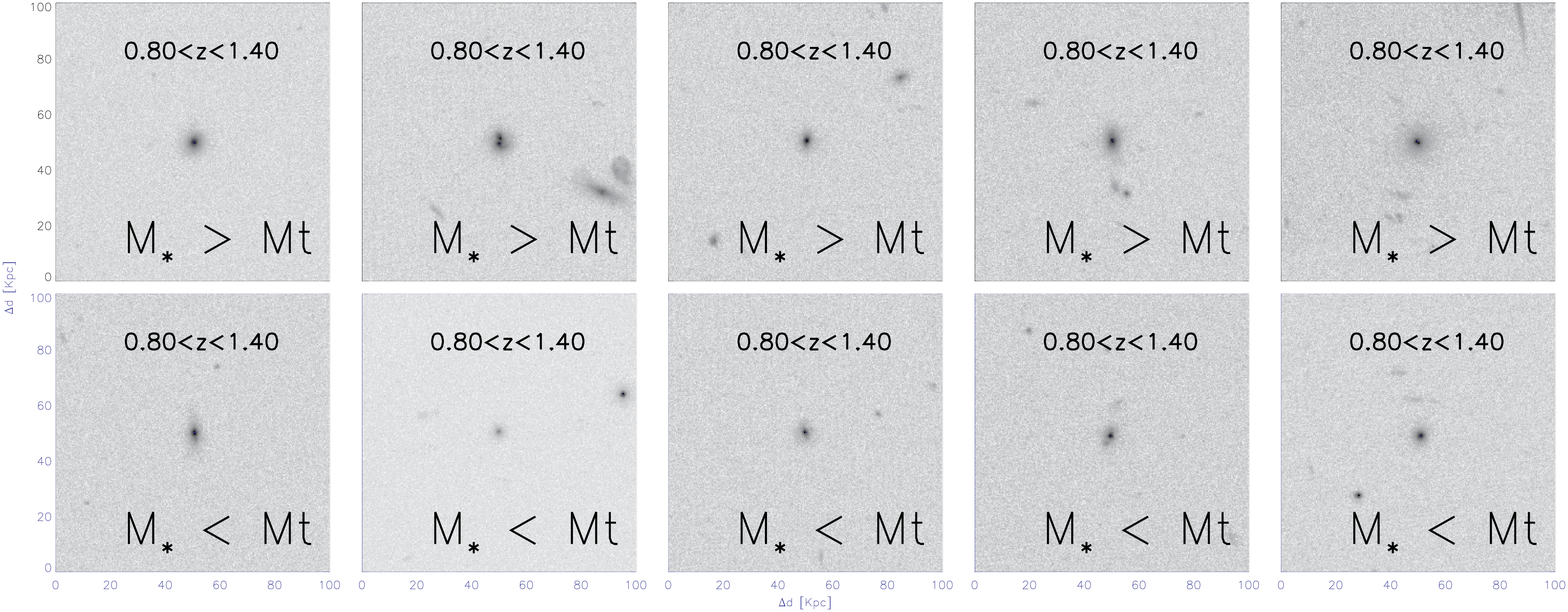}}
 \caption{$\sim100Kpc\times100Kpc$ I-band ACS stamps of blue-sequence E/S0 galaxies in the three redshift bins. For illustration, we show galaxies with stellar masses greater and lower than the threshold mass at a given z.} 
 \label{fig:morpho_stamps1}
 \end{figure*}

Given that the relative abundance of blue E/S0 depends on mass, it is interesting  to see wether there are  some differences within the
blue                                                         early-type
population in terms of morphology. Figure~\ref{fig:morpho_stamps1} shows some  example  stamps  of  blue
early-type galaxies in the 3 considered redshift  bins split into 2 mass bins ($M_*/M_\odot>M_t$ and $M_*/M_\odot<M_t$) . Most of the blue  E/S0  galaxies  present regular morphologies with high central light concentration as shown in section~\S~\ref{sec:def}. As a general trend, massive galaxies tend to be regular but with a diffuse \emph{disky} component in the outer-parts. Less massive galaxies are more compact and it is more difficult to detect internal structures. At   $z=0$,
\cite{kannappan09}  detected in  fact that  less massive  galaxies were
more disturbed than the massive counterparts. They interpreted this as
a signature  of a  different regime  of the  blue early-type
population. In our case it is harder to identify this kind of details,
specially  in  the  higher  redshift  bin,  despite  the  high-angular
resolution delivered by HST, because of cosmological dimming effects. In the lower redshift bin however (top panel of figure~\ref{fig:morpho_stamps1}) it seems that low mass blue E/S0 tend to have a higher concentration of small companions around than more massive galaxies which seem to be more isolated. Of course, we would need the redshifts of these faint companions to confirm this trend, which are not available for the moment.

\subsection{Size}

\begin{figure*} 
 \centering 
 {\includegraphics[width=17cm]{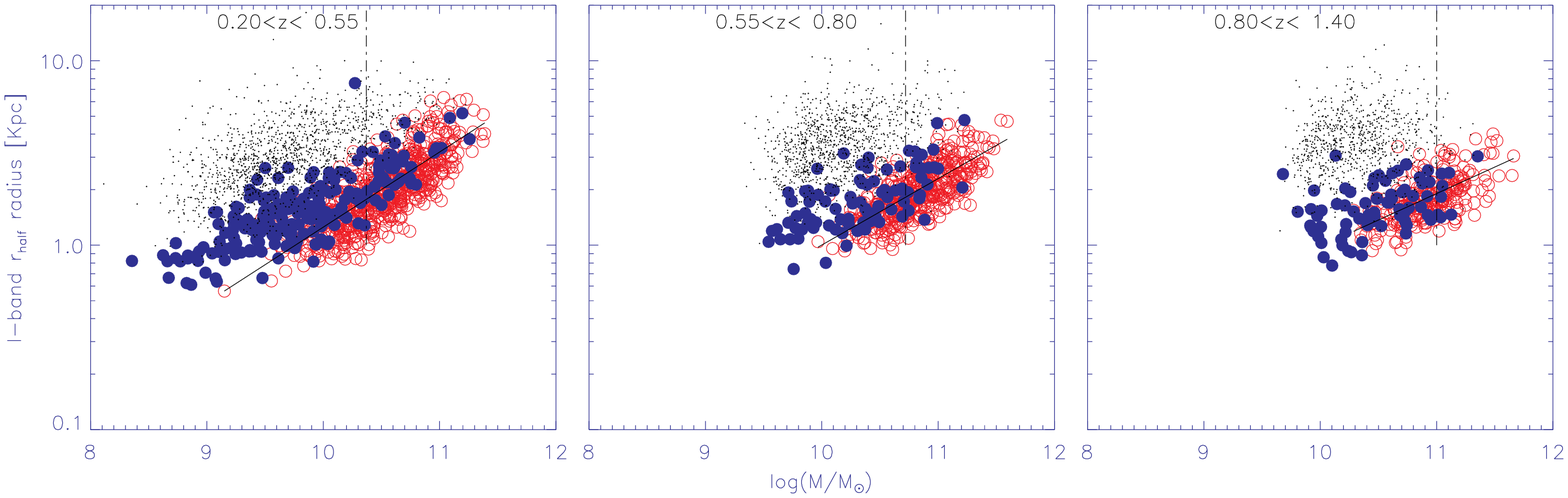}}
 \caption{Radius-stellar mass relation at different redshifts. Red circles are red E/S0 galaxies, black points are blue late-type systems and filled blue circles are blue early-type galaxies. Vertical dashed lines show the threshold mass at different redshifts as computed in \S~\ref{sec:numbers}. Solid line indicates the best fit line for red E/S0 galaxies.  } 
 \label{fig:mass_radius}
 \end{figure*}

In figure~\ref{fig:mass_radius}, we plot the $M_{*}-radius$ relation for blue E/S0 galaxies compared to red E/S0 and blue late-type galaxies. The $r_{half}$ radii are computed in the I-band ACS images and are defined as the radius containing half of the flux of the galaxy. Globally, blue E/S0 galaxies are closer to red E/S0 than to late-type systems in
the   $M_{*}-radius$  relation  (Fig.~\ref{fig:mass_radius})   at  all
redshifts. We
detect however  some tendencies with  the stellar mass.  It seems that, below a given mass, blue E/S0 galaxies tend to deviate from red E/S0 in the $M_{*}-radius$ plane and lie closer to normal star-forming late-type galaxies. In order to quantify this deviation we perform a linear fit to the $M_{*}-radius$ of red E/S0s and compute the median distance of blue E/S0 galaxies to the best fit line as a function of mass.  It rises from $\Delta r_{half}/\overline{r_{half}}\sim0.1$ Kpc for $log(M_*/M_\odot)>10.5$ to  $\Delta r_{half}/\overline{r_{half}}\sim0.5$ Kpc for $log(M_*/M_\odot)<10.5$. The transition mass does not seem to change significantly with redshift. 

%To quantify this scatter increase we fit the $M_{*}-radius$ relation for red E/S0 galaxies and compute the scatter of blue early-type galaxies to the best fit line below $M_t$ and above $M_t$. We obtain that the scatter 

%\begin{comment}
%\begin{figure*} 
 %\centering 
  %\resizebox{\hsize}{!}{\includegraphics{blueE_stamps_lowZ_highD.ps}}
  %\caption{$300kpc\times300kpc$ stamps of blue-sequence E/S0 galaxies in the low redshift range ($0.2<z<0.55$) lying in high density environments ($D>0.1$).  } 
 %\label{fig:morpho_stamps1}
 %\end{figure*}
% \end{comment}

\subsection{Star formation}
Star  formation  rates are  estimated  with  the  equivalent width  of
[OII]$\lambda$3727$\rm{\AA}$ line measured  on  the   publicly  available
zCOSMOS  DR2  spectra \citep{Lilly09}.  We  measure equivalent  widths
using our own code and transform  the measured values to SFR using the
relation             derived             in            \cite{guzman97}
(Eq.~\ref{eq:SFR}).  

\begin{equation}
SFR(M_\odot yr^{-1})\sim2.5\times10^{-12}\times10^{-0.4(M_B-M_{B\odot})}EW_{[OII]}
\label{eq:SFR}
\end{equation}

\begin{figure*} 
 \centering 
{\includegraphics[width=17cm]{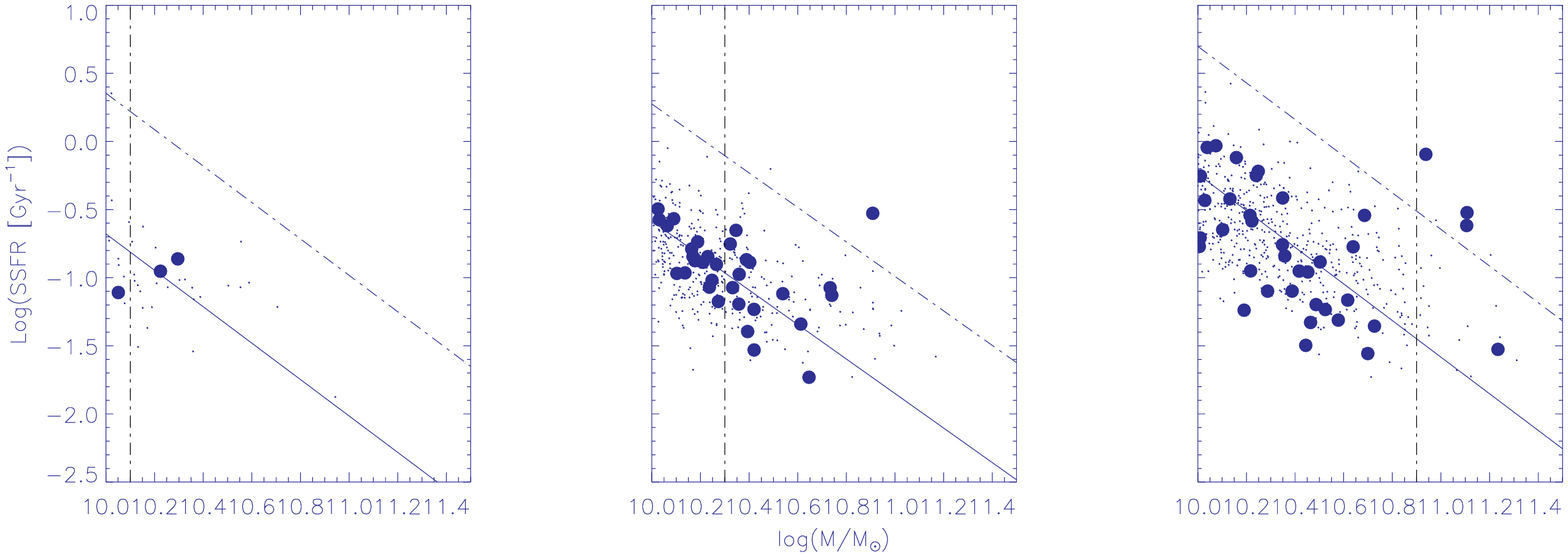}}
  \caption{Specific star formation rate as a function of stellar mass in 3 redshift bins. Black dots are normal blue late-type galaxies and blue filled dots are blue early-type galaxies. Vertical dashed lines indicate the threshold mass. Solid line is the best fit to the spiral star-forming population and dashed line is the 3-sigma limit.} 
 \label{fig:SFR}
 \end{figure*}

Figure~\ref{fig:SFR}   shows  the  specific  star
formation rate (SSFR) for  \emph{normal} late-type galaxies compared to
the blue  early-type ones  at different redshifts.  We do  not show the results for red early-type galaxies since very few presented [OII]
emission  lines. Notice  also that  most  of the  galaxies located at
$0.2<z<0.55$ do not  have measurements of their SSFR  due to the [OII]
emission line lies out of zCOSMOS spectral range ($5000\AA$-$9000\AA$).

Globally  the  measured  SSFR   is  comparable  for  both  populations
suggesting that  blue early-type galaxies might have  a disk component
(either formed  or in formation) which  is accounting for  the bulk of
the star  formation. Only for  very massive blue-early  type galaxies,
above  the  threshold  mass,  we  seem to  detect  outliers  in  this
relation. These galaxies have indeed high specific star formation rates ($\sim 10^{-0.5}$ $Gyr^{-1}$) which are characteristic of a post-merger phase. 

% Numbers are small since there are very few blue E/S) in this mass regime and it therefore becomes very difficult to conclude safely. However if this effect was real it will agree with the scenario in which more massive galaxies are mostly merger remnants suffering a violent episode of star-formation which will migrate to the red-sequence while less massive galaxies will become today spirals. 

\section{Discussion}
\label{sec:discussion}
\subsection{Are blue E/S0 blue compact galaxies?}

\begin{figure} 
 \centering 
  \resizebox{\hsize}{!}{\includegraphics{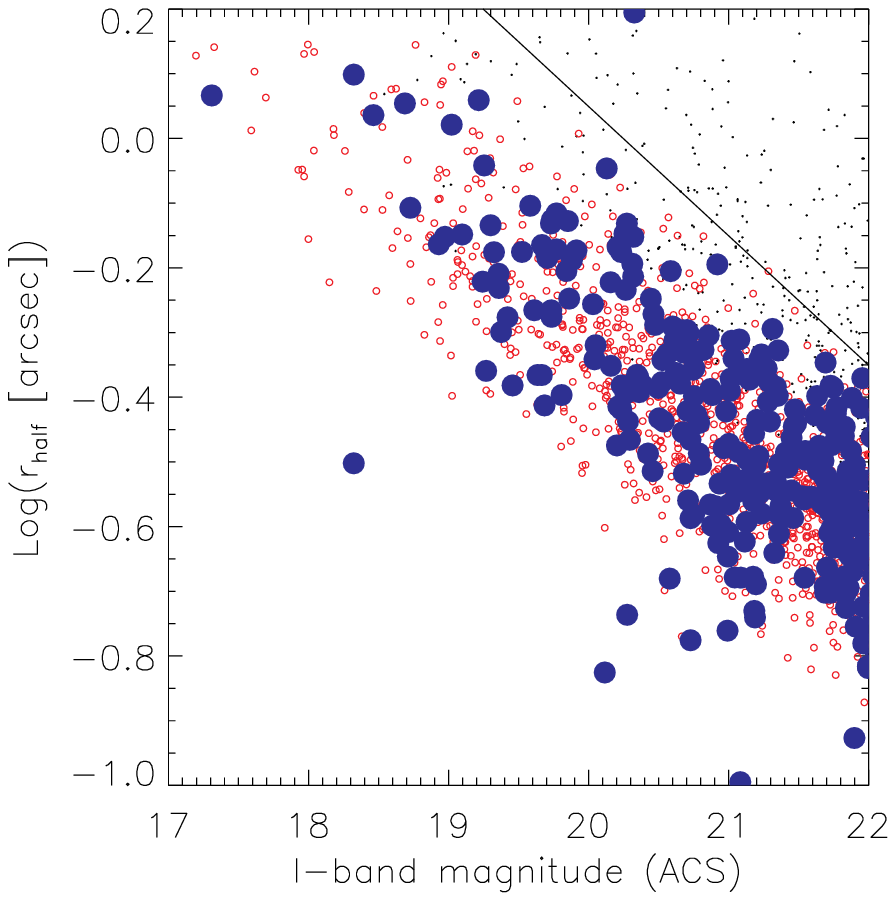}}
 \caption{I-band magnitude versus half light radius. Blue filled circles are our blue early-type galaxies, empty red circles are red E/S0 and black dots are blue late-type galaxies. The solid line shows the limit used in \cite{phillips97} to define compact objects. } 
 \label{fig:i_rad}
 \end{figure}

There  is  a well-known  family  of  galaxies  called blue  compact
galaxies. These galaxies were defined by \cite{guzman97} as
objects  presenting a  high surface  brightness within  the half-light
radius.  The spectroscopic  study \citep{phillips97}  revealed  that an
important   fraction   of  these   objects   present  emission   lines
characteristic  of  on-going  star   formation  and  hence  have  blue
integrated colors.  Our definition of  blue E/S0 is different since it
is based  on the  morphology, however we expect these galaxies to have high central light concentrations.  One interesting question is therefore: are our  galaxies the same  blue compact
galaxies?  In  Figure~\ref{fig:i_rad} we  show  the  I band  magnitude
versus the half light radius both computed in the ACS images. Our blue
early-type galaxies indeed fall in the region of blue compact galaxies
defined by \cite{phillips97}.

\begin{figure} 
 \centering 
  \resizebox{\hsize}{!}{\includegraphics{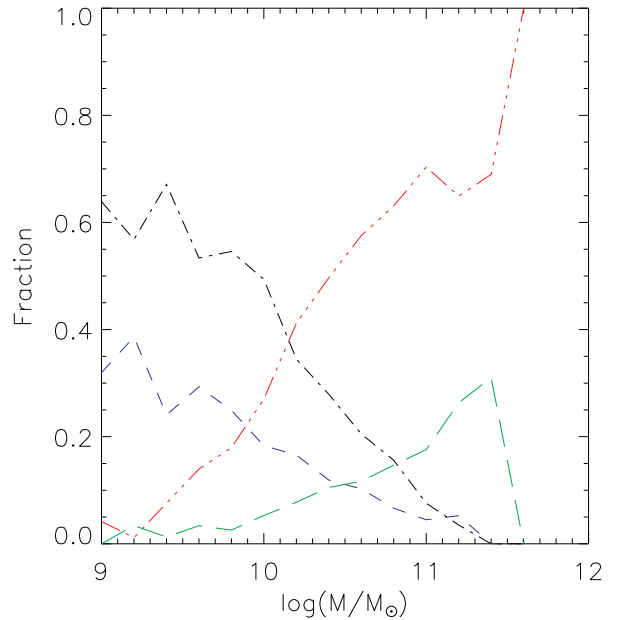}}
 \caption{Morphological mixing of compact galaxies as a function of the stellar mass. Red line are passive early-type galaxies, blue line are blue early-type galaxies, black line blue late-type galaxies and green line red late-type galaxies.  } 
 \label{fig:compact_frac}
 \end{figure}

There is however  an important difference with respect  to their work:
while  \cite{phillips97} claimed  that an  important fraction  of these
compact  galaxies present  blue colors,  we  find that  the region  is
mostly dominated by passive early-type galaxies. This difference might
be  explained by  incompleteness  effects. They indeed used a small and deep ($I<24$) flux  limited
sample     and    since    red     galaxies    are     lost    earlier
(Fig.~\ref{fig:mass_z}), the fraction of  blue galaxies in this region
might be  severely overestimated. In  figure~\ref{fig:compact_frac} we
show the morphological  mixing of compact galaxies in  our sample as a
function of  the stellar mass.  Blue galaxies start to  dominate among
compact   galaxies  at  low   masses  ($M_*<10^{10}   M_{\odot}$)  where
incompleteness effects in the  red  population become important.  We
argue therefore, as stated also in previous works (e.g. \citealp{Ilbert06}) that the high  fraction of blue compact galaxies found
in    \cite{phillips97}    might    be    biased   because    of    the
incompleteness.  Indeed,  their fields  were deeper  and
covered a  smaller area  than ours, so  their sample was  dominated by
low-mass compact galaxies, while we  focus here in the massive tail of
the distribution.

Moreover, \cite{guzman97} concluded  that  most  of their  blue  compact
galaxies at  high redshift  (about 60$\%$) were  similar to  local HII
regions. We  have inspected the spectral energy  distribution (SED) of
our  blue E/S0 galaxies  and  conclude that  they  are not  similar to  HII
galaxies.  Their SEDs  are  similar to  normal  starbust galaxies.  We
conclude that our  blue E/S0 galaxies indeed are  compact galaxies but
are  not the same  population of  objects as  the studied  in these previous
works.

\subsection{Characteristic masses}

In section~\ref{sec:numbers} we have  shown that there  is no evolution of  the blue
E/S0  shutdown mass ($M_s\sim3\times10^{11}$).  This shutdown  mass is
indeed similar  to the  mass found for  blue late-type  galaxies. This
implies that there are no  blue galaxies above $M_s$ from $z\sim1$. It
suggests that  this value  is not linked  to the blue  E/S0 population
only but it is a general property of blue galaxies. This has also been
found  in   other  recent   works  \citep{cattaneo06}  and   has  been
interpreted as  a critical mass  above which the blue  sequence cannot
exist  because the  gas physics  is  different. Only  haloes below  a
critical shock-heating  mass have  gas supply by  cold flows  and form
stars. In contrast, cooling  and star formation are shut-down abruptly
above  this  mass. \cite{dekel06}  proposed  that  this critical  mass
should  be constant  from  $z\sim2$  which is  in  agreement with  our
findings.

The behavior  of the threshold mass  is different. We  have measured a
decrease  of  $\sim1$  dex  from $z\sim1$  to  $z\sim0.2$.   Recently
\cite{kannappan09} measured this threshold  mass at $z\sim0$ and found a
value of $\sim8\times10^9 M_\odot$. This gives a total decrease of 1.4
dex from  z$\sim1$ to  $z\sim0$. \cite{kannappan09} also  suggested that
the  threshold mass  indicates  a turnover  in  the dominant  physical
mechanisms taking place  in blue E/S0 and that it  should be linked to
the  bimodality  mass.  Blue  E/S0  galaxies  less  massive  than  the
threshold  mass at  $z\sim0$ are  mainly rebuilding  disks  while more
massive galaxies are  galaxies in a post-merger phase.  The value they
found is $1$  dex smaller than the crossover  mass between the passive
early-type population and the star-forming late-type one in the nearby
universe (bimodality mass) \citep{Kauffmann06} and they suggest a link
between them.

Since  the bimodality  mass rises  at high  redshift  (i.e. downsizing
effect,  \citealp{bundy05}),   \cite{kannappan09}  predicted  a  similar
behavior  for the  blue E/S0  threshold mass,  remaining $\sim  1$ dex
below  the bimodality mass  at a  given z.  As shown  in \S~\ref{sec:numbers}
$M_b$ evolves  becoming higher at higher redshifts  and reflecting the
fact that  the building-up of  the red-sequence is preformed  first at
high masses.  However, the measured  values are fully  compatible with
the values measured for the  threshold mass and we consequently do not
find this expected difference between both characteristic masses. This
suggests that  both values  are measuring the  same process,  i.e. the
build-up of the RS. In fact, blue E/S0 start to be important in number
when  the RS galaxies  are no  longer dominant  among the  blue active
population.  We  argue therefore  that  the  bimodality  mass and  the
threshold mass are  in fact two different quantifications  of the same
evidence. In other words,  blue early-type galaxies have a significant
importance only for masses at which the RS is not yet built.

\subsection{Disk rebuilding or merger remnants?}

The question is  now: is this bimodality-threshold mass  or any other mass linked to the
nature of blue E/S0 galaxies as suggested by \cite{kannappan09} at low
z? When  looking at the SSFR  at different masses (Fig.~\ref{fig:SFR}) we  observe that all
the  outliers in  the linear  SSFR-$M_*$  relation  are located
above $log(M_*/M_\odot)\sim10.8$ at all redshifts which is bigger than the threshold mass ($M_t$):  above this mass, blue
E/S0  tend  to  have  higher  specific star  formation  rates  ($>10^{-0.5}$
$Gyr^{-1}$). By definition, the number of galaxies above $M_t$ is small
so it  becomes difficult  to establish robust  conclusions but  we can
affirm that  all the outliers  fall above $log(M_*/M_\odot)\sim10.8$. This  result suggests
indeed that there  might be a change in  the physical processes taking
place in  blue E/S0 depending on the  mass. This high SSFR  is in fact
characterisitc  of strong  starburst  (e.g. \citealp{springel05})  and
these objects could be galaxies experiencing a violent episode of star
formation  in  a  post  merger  phase. Their  normal  evolution  would
therefore be a migration to  the RS after consumption of the available
gas.   This    later   affirmation   is   also    supported   by   the
$r_{half}$--$M_{*}$ relation in  which we  can see  that massive
blue E/S0 galaxies tend to fall  within the same region as passive red
E/S0s at all redshifts (Fig.~\ref{fig:mass_radius}). The visual inspection of these massive galaxies (Fig.~\ref{fig:blueE_stamps}  and~\ref{fig:morpho_stamps1}) does not reveal important perturbations up to our resolution limit, suggesting that these galaxies have already acquired their early-type morphology and are ready to migrate into the RS. Moreover, they seem to lie in low density environments or at least not close to comparable mass companions (Fig.~\ref{fig:blueE_stamps}  and~\ref{fig:morpho_stamps1}) which seems to indicate that they will probably not experience a major merger event in the next Gyrs. 

%These galaxies are morphologically similar to normal E/S0 galaxies but with a very high star formation rate. 

\begin{figure} 
 \centering 
  \resizebox{\hsize}{!}{\includegraphics{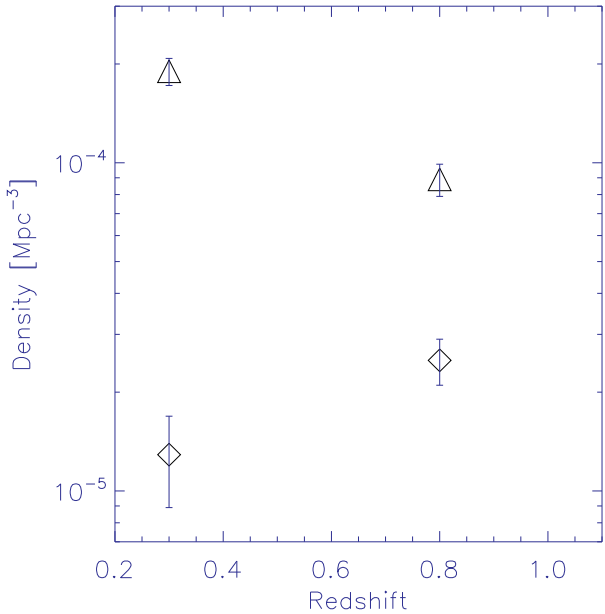}}
 \caption{Evolution of number density of blue E/S0 in two mass bins. Diamonds are blue E/S0 with $log(M/M_\odot)>10.8$ and triangles are blue E/S0 with $10<log(M/M_\odot)<10.8$. } 
 \label{fig:blueE_density}
 \end{figure}

According  to numerical  simulations  of mergers  of massive  gas-rich
galaxies, the typical time scale to move from the blue cloud
to   the  RS   after   a   major  merger   event   is  $\sim2-3$   Gyr
\citep{springel05}.  Consequently,  if  the  most  massive  blue  E/S0
galaxies were indeed galaxies in  a post-merger phase, we could expect
a decrease of the density of these objects in this time interval since
we  expect   a  decrease   in  the  merger   rate  with   cosmic  time
(e.g. \citealp{conselice09, conselice08, leFevre00}). In order to check this, we have computed
the number density of blue E/S0  galaxies in two redshift bins of 3Gyr
each  ($0.2<z<0.55$  and  $0.55<z<1.4$)   for  two  mass  regimes.  In
Figure~\ref{fig:blueE_density} we  see that the comoving number  density of the
most massive blue E/S0 galaxies ($log(M/M_\odot)>10.8$) decreases from
$z\sim0.8$ to $z\sim0.1$ of a factor $\sim2$. This suggests that these
galaxies have time to move to the RS in less than $3$ Gyr, as expected
from major merger remnants. 

At lower masses ($log(M_*/M_\odot)<10.8$)  however, the SSFR of blue E/S0 is  similar to the one
of   star-forming  late  type   galaxies  (Fig.~\ref{fig:SFR})
suggesting   that  these   galaxies   are  more   similar  to   normal
disks. Moreover,  if we look at the  $r_{half}$--$M_{*}$ relation (Fig.~\ref{fig:mass_radius})
we do  see that these galaxies  start to deviate from  the relation of
passive  E/S0  and  lie  between  the  late-type  and  the  early-type
zone. They are consequently smaller  than normal rotating disks at the
same  redshift but  larger than spheroids. This  suggests  that these
galaxies might evolve differently than their massive counter-parts. In
any case, the  main driver of their evolution  should have time scales
larger than $3$ Gyr since we do not observe a decrease in their number
density in  this time interval  (see Fig~\ref{fig:blueE_density}). Our
guess is that 1) they can evolve by fading until becoming normal disks, 2) they  can be  rebuilding  disks from
surrounding gas \citep{governato07, robertson08} 
as  suggested by  \cite{hammer01} or 3) they can have built a bulge component through minor mergers and be in a reddening phase or \emph{morphological quenching} phase as proposed by \cite{martig09}. The last two hypothesis require gas accretion: in this case where is the gas coming from? It does not seem that the presence of gas is a consequence of a major merger event since we do not detect any outlier in the SSFR-$M_*$ relation and disk rebuilding after a major merger event can be done in a relatively short time-scale \citep{hammer09}. It is more likely that these galaxies are the result of minor mergers which triggered the star formation in the central parts \citep{aguerri01, elichemoral06} and built a bulge component \citep{martig09}. This process would also imply a disk growth since most of the satellites mass is expected to fall in the outer parts. Even if the analysis of their environment reveals that the majority of them do not lie in dense environments up our completeness limit, all blue E/S0 galaxies lying in relatively high environments ($D3>0.1$) have low masses and we cannot reject the possibility that they might have gas-rich low-mass companions as we seem to detect in the ACS images. If this scenario revealed to be true, according to the numerical simulations performed by \cite{martig09}, these galaxies then will redden in $\sim4-5$ Gyr between $z\sim1$ and $z\sim0.2$ through what they called a \emph{morphological quenching} phase. 

%Secular evolution can also contribute to the evolution of these galaxies, some studies have indeed shown that there is cold gas surrounding galaxies (e.g. \citealp{combes97, combes99}) which can contribute to the disk growth. \\

%In any case, according to these results, it seems difficult that they might become passive early-type galaxies without a merging process.  \\

\begin{figure*} 
 \centering 
{\includegraphics[width=17cm]{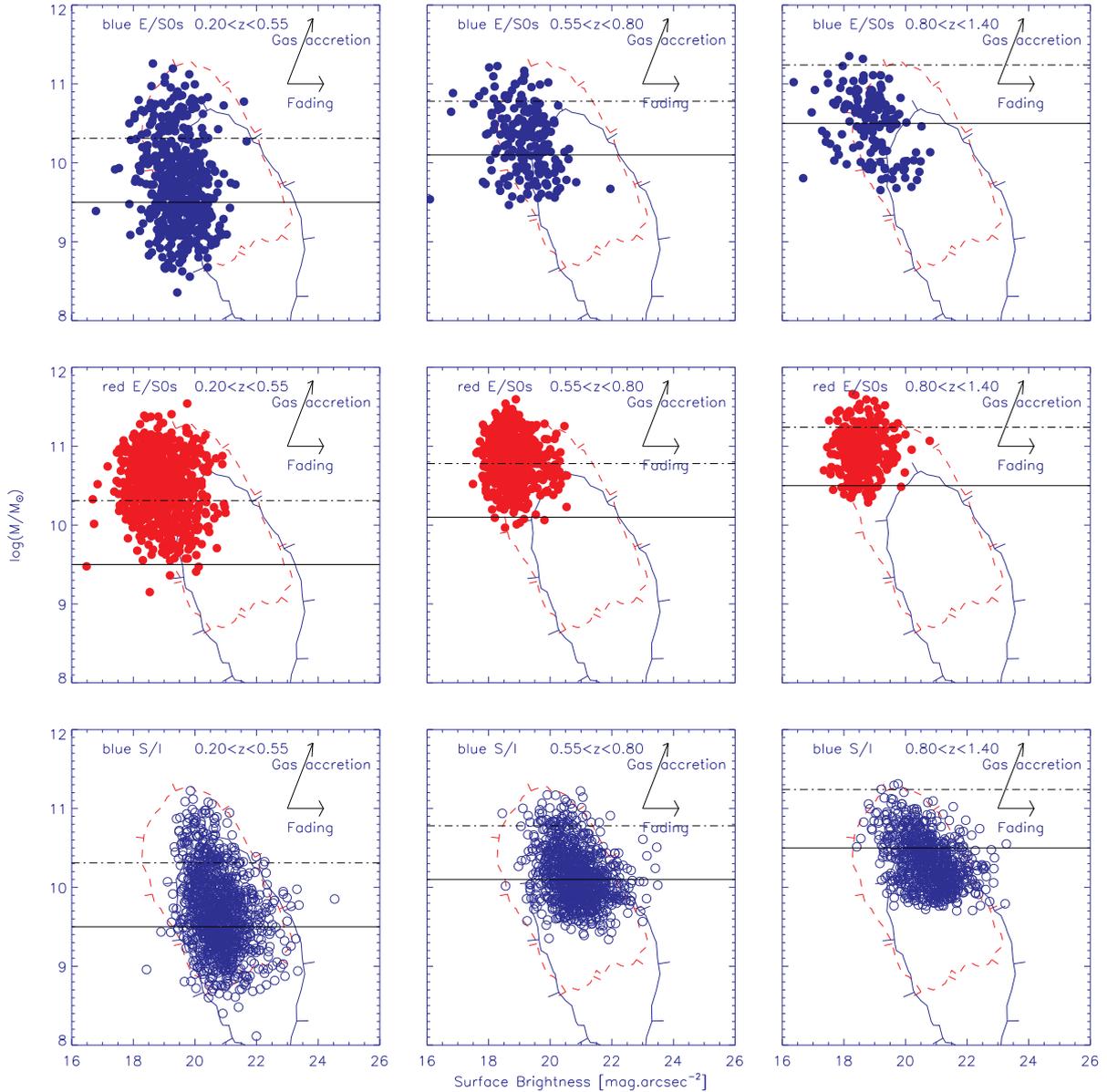}}
 \caption{R-band rest-frame surface brightness vs. stellar mass for our sample compared with the SDSS DR4 sample. Each column shows a different redshift range, i.e. from left to right: $0.2<z<0.55$, $0.55<z<0.8$ and $0.8<z<1.4$. Contours show in all sub-figures the location (20 galaxies iso-contour) of local galaxies (red dashed contours are early-type galaxies and blue solid contours are late-type galaxies). Over plotted are galaxies in our sample, top: blue E/S0, middle red E/S0 and bottom blue late-type systems. Dashed horizontal lines indicate the threshold mass and the solid line are our completeness limit. Arrows show typical movements in the plane caused by fading or disk rebuilding activity. } 
 \label{fig:sb_mass}
 \end{figure*}

In  order to  investigate  further  the future  of  these objects and in particular the fading scenario for the less massive galaxies,  we
compare  in figure~\ref{fig:sb_mass}  the surface  brightness  -- mass
relation  of our objects  with local galaxies  from the SDSS DR4
\citep{Adelman-McCarthy06} and more precisely from the \emph{low redshift} catalog described in \cite{Blanton05}.  We define a local galaxy to be early-type or late-type based on the Sersic index computed by \cite{Blanton05}. This definition is slightly different than the one used for the high-redshift sample but we do not expect the differences to be significant since there is a very tight relation between the Sersic index and the visual morphology at low redshift. The surface brightness is computed in the r rest-frame band in both samples to avoid wavelength dependent effects. Massive blue  early-type galaxies  at high
redshift overlap  with the zone of local  passive early-type galaxies,
indicating that they will probably transform into this class. However,
less  massive galaxies  lie in  a different  zone which  has  no local
counter   part,   suggesting  that   they   are  experiencing   strong
evolution. In order to reach the local star-forming region a fading of
$\sim2$ magnitudes in  the r band is required.   Nevertheless, even if
E/S0  blue  galaxies could  suffer  this  fading,  the resulting  spiral
galaxies would be smaller than the observed normal spirals galaxies in
the  nearby  Universe.  As a matter of fact, we  do  not expect  a  large
increment of  the scale for a  late-type galaxy evolving  just by pure
passive   evolution  (see   \citealp{trujillo04,   brook06,  kanwar08,
  azzollini08}). In contrast the (re)building of the outermost regions of
the galaxies by  the accretion of gas would increase  the mass and size
of  the galaxies moving  E/S0 blue objects to  the positions occupied by
normal local galaxies in the surface brightness-mass plane. \\

%The nature of blue E/S0 indeed seems to depend on the stellar mass. However, we find that the turn-over mass is close to $log(M_*/M_\odot)\sim10.8$ at all redshifts and does not seem to evolve significantly from $z\sim1$ in contrast with the threshold mass. Assuming therefore that all blue E/S0 galaxies with masses $9.8<log(M_*/M_\odot)<10.8$ are progenitors of nowadays spirals, we can even go further and estimate that the fraction of \emph{disks} experiencing a strong evolution from $z\sim1$ is at least $\sim11\%$. 

\section{Summary and conclusions}
\label{sec:conclusions}
We have studied some of the properties of a population of 210 blue early-type galaxies with $M_*/M_{\odot}>10^{10}$ from $z\sim1$ in the COSMOS field. Our sample is complete up to $M_*/M_{\odot}\sim10^{10}$ in the lower redshift bin and the completeness limit rises to $M_*/M_{\odot}\sim5\times10^{10}$ at $z\sim1$. Spectroscopic redshifts come from the zCOSMOS 10k release. Morphologies are determined in the ACS images using our tested and validated code \textsc{galSVM} and the spectral classification is performed using a best-fitting technique with 38 templates on the primary photometric COSMOS catalogues. We define a blue E/S0 galaxy as an object presenting an early-type morphology and which SED best fits with  a blue template. \\

The main properties of these galaxies summarize as follows:

\begin{itemize}

\item Globally blue E/S0 galaxies represent $\sim5\%$ of the whole sample of early-type galaxies with $M_*/M_{\odot}>10^{10}$ from $z\sim1.4$. Nevertheless the abundance strongly depends on the considered stellar mass. Below a threshold mass ($M_t$), they become more abundant and represent about $20\%$ of the red early-type population. This threshold mass evolves with redshift from $M_t\sim11.1$ at $z\sim1$ to $M_t\sim10.1$ at $z\sim0.2$. This evolution matches the evolution of the bimodality mass (crossover mass between late-type blue population and the early-type red population). Both masses are therefore a measure of the build-up of the RS. In contrast, the shutdown mass (mass above which blue galaxies dissappear) does not evolve. This could be related with differences in the gas physics above a critical mass. 

\item Blue E/S0 galaxies fall in the same region in the magnitude-radius space than blue compact galaxies defined by \cite{guzman97}. However it is not the same population since we focus here in the massive tail of the distribution which is dominated by passive red E/S0 galaxies. 

 \item The visual inspection of these galaxies does not reveal important perturbations up to out resolution limit. On the contrary, galaxies seem to have a well-defined shape with high-central light concentration. Low mass galaxies seem to have more small companions than the massive counter-parts, however we do not have spectroscopic information to confirm this trend.

\item Massive blue E/S0 ($log{M_*/M_\odot}>10.5$) galaxies tend to have similar sizes than E/S0 galaxies. At lower masses however they tend to lie between spirals and spheroids. 

\item The specific star formation rate measured with the [OII]3727 line and stellar masses, shows that blue E/S0 galaxies have similar SSFR values than normal star-forming spirals. The outliers of this relation are galaxies more massive than $log(M_*/M_\odot)\sim10.8$. These outliers present SSFR characteristic of post-merger galaxies ($\sim 10^{2.5}$ $Gyr^{-1}$).   

\begin{comment}
\item We find that blue E/S0 galaxies are located in similar environments than normal spirals. They are mostly isolated galaxies up to our completeness limit, i.e. they do not seem to have companions of comparable mass. However, all the objects lying in relatively high environments have $log(M_*/M_\odot)<10.8$.
\end{comment}

 \end{itemize}

These results seem to point out that these galaxies have different natures depending on their stellar mass. Massive objects ($log(M_*/M_\odot)>10.8$), are probably post mergers galaxies migrating to the red-sequence while less massive galaxies are more likely progenitors of todays late-type galaxies which are accreting gas. The migration of the massive galaxies is faster than $3$ Gyr as expected by numerical simulations of merger remnants. Indeed, the evolution of the number density computed in redshift bins of equal time show that the most massive E/S0 galaxies evolved in typical time-scales smaller than 3 Gyr. In contrast the evolution of blue early-type galaxies with smaller masses is produced in larger time-scales. This low-mass galaxies might be the result of minor-mergers with surrounding satellites which triggered the central star-formation and built-up a bulge component, and therefore appear with an early-type shape. \\
The confirmation of these hypothesis requires a detailed analysis of the internal kynematics of these objects to see if the physics is indeed different. In particular, 3D spectroscopy should be useful to get more details about their nature.

\begin{acknowledgements}Based on zCOSMOS observations carried out using the Very Large Telescope at the ESO Paranal Observatory under Programme ID: LP175.A-0839. The Millennium Simulation databases used in this paper and the web application providing online access to them were constructed as part of the activities of the German Astrophysical Virtual Observatory. The authors would like to thank ESO for financial support. 
\end{acknowledgements}

\bibliographystyle{aa}
\bibliography{biblio}

\end{document}